\documentclass[twocolumn]{aastex631}

\usepackage{longtable}
\usepackage[flushleft]{threeparttable}
\usepackage{tabularx}
\usepackage{multirow}
\usepackage{graphicx}
\usepackage{amsmath,amssymb}
\usepackage{color}
\usepackage{units}
\usepackage{epstopdf}
\usepackage{hyperref}
\usepackage{multirow}
\usepackage{url}
\usepackage{subfigure}
\usepackage{rotating}
\usepackage{enumitem}\setlist[description]{font=\textendash\enskip\scshape\bfseries}

\newcommand{\beq}{\begin{equation}}
\newcommand{\eeq}{\end{equation}}
\newcommand{\bdm}{\begin{displaymath}}
\newcommand{\edm}{\end{displaymath}}
\newcommand{\T}[1]{\tilde{#1}}

\definecolor{Gray}{gray}{0.9}
\definecolor{orange}{rgb}{0.9,0.5,0}

\newcommand{\blue}[1]{\textcolor{black}{#1}}

\begin{document}

\title{Foraging with MUSHROOMS: A Mixed-Integer Linear Programming Scheduler for Multimessenger Target of Opportunity Searches with the Zwicky Transient Facility}

\author[0000-0002-3155-0385]{B. Parazin}
\affil{College of Science, Northeastern University, Boston, Massachusetts 02115, USA}
\affil{School of Physics and Astronomy, University of Minnesota, Minneapolis, Minnesota 55455, USA}

\author[0000-0002-8262-2924]{Michael W. Coughlin}
\affil{School of Physics and Astronomy, University of Minnesota, Minneapolis, Minnesota 55455, USA}

\author[0000-0001-9898-5597]{Leo P. Singer}
\affil{Astrophysics Science Division, NASA Goddard Space Flight Center, Code 661, Greenbelt, MD 20771, USA}
\affil{Joint Space-Science Institute, University of Maryland, College Park, MD 20742, USA}

\author[0000-0002-7672-0480]{Vaidehi Gupta}
\affil{Indian Institute of Technology, Kharagpur, West Bengal 721302, India}

\author[0000-0003-3768-7515]{Shreya Anand}
\affil{Division of Physics, Mathematics and Astronomy, California Institute of Technology, Pasadena, CA 91125, USA}

\begin{abstract}
Electromagnetic follow-up of gravitational wave detections is very resource intensive, taking up hours of limited observation time on dozens of telescopes. Creating more efficient schedules for follow-up will lead to a commensurate increase in counterpart location efficiency without using more telescope time. Widely used in operations research and telescope scheduling, mixed integer linear programming (MILP) is a strong candidate to produce these higher-efficiency schedules, as it can make use of powerful commercial solvers that find globally optimal solutions to provided problems . We detail a new target of opportunity scheduling algorithm designed with Zwicky Transient Facility in mind that uses mixed integer linear programming. We compare its performance to \texttt{gwemopt}, the tuned heuristic scheduler used by the Zwicky Transient Facility and other facilities during the third LIGO-Virgo gravitational wave observing run. This new algorithm uses variable-length observing blocks to enforce cadence requirements and ensure field observability, along with having a secondary optimization step to minimize slew time. \blue{We show that by employing a hybrid method utilizing both this scheduler and \texttt{gwemopt}, the previous scheduler used, in concert, we can achieve an average improvement in detection efficiency of 3\%-11\% over \texttt{gwemopt} alone} for a simulated binary neutron star merger data set consistent with LIGO-Virgo's third observing run, highlighting the potential of mixed integer target of opportunity schedulers for future multimessenger follow-up surveys.

\end{abstract}



\section{Introduction}
\label{section:intro}

The detection of GW170817 \citep{AbEA2017b} in August 2017 signaled the beginning of a new era of multimessenger astronomy, promising advances in r-process nucleosynthesis \citep[e.g.][]{2017ApJ...848L..19C,2017Sci...358.1556C,2017ApJ...848L..17C,2017Natur.551...67P}, the neutron star equation of state 
\citep[e.g.][]{2018PhRvL.121p1101A,2018ApJ...852L..29R, 2019MNRAS.489L..91C,2018MNRAS.480.3871C, 2020MNRAS.492..863C,2020Sci...370.1450D}, and the value of the Hubble constant \citep[e.g.][]{2017Natur.551...85A,2019NatAs...3..940H,2020Sci...370.1450D}.
This was thanks to the detection of GW170817 \citep{PhysRevLett.119.161101} and its electromagnetic counterparts: a kilonova (ultraviolet/optical/near-IR emission generated by the radioactive decay of r-process elements) \citep[e.g.][]{2017kilonova,2017Sci...358.1565E,2017Sci...358.1583K,2017Natur.551...67P,2017Sci...358.1574S,2017Natur.551...75S}, a short gamma ray burst \citep[e.g.][]{2017GRB}, and an afterglow \citep[e.g.][]{2017afterglow,2017Natur.551...71T}.
However, since then, no further electromagnetic counterparts to gravitational-wave detections have been confirmed, despite several other detected binary neutron star and neutron star black hole mergers during LIGO-Virgo's third observing run \citep{O3CBC1}. This can mostly be explained by the localization areas of neutron star containing mergers being much larger than expected, \citep[e.g.][]{CoDi2020,PeSi2021}, making efficient observation planning all the more important. With, on average, a much larger area than previously thought to search, there are many more choices for potential schedules, but it will take an optimal scheduler to maximize scientific output.

Fundamentally, telescope scheduling software determines which fields to observe in what order, subject to environmental and programmatic constraints. In the case of the follow-up of large sky localizations produced by gravitational-wave \citep[e.g.][]{CoAh2019b,AnCo2020} or gamma-ray burst \citep[e.g.][]{CoAh2019,AhSi2021} events with wide field of view surveys such as the Zwicky Transient Facility \citep{Bellm2018,Graham2018,DeSm2018,MaLa2018}, the goal is usually to maximize an objective function, which is typically taken to be the integral of the probability skymap over the combined footprint of all the observations, although other choices are possible \citep{CoTo2018}.
These observations should also be completed in the minimal amount of time, as many different science programs time share on the same telescope, and therefore any time saved can be utilized by other science programs \citep{Bellm:19:ZTFScheduler}.

While, in principle, this could be done manually, schedules designed this way are labor-intensive and sub-optimal, and it is unclear how the heuristics translate to survey effectiveness.
Another common approach is the use of ``Greedy'' algorithms, which compute a metric or score for each possible target, select the target with the current highest value, observe it, and then repeat the process.
This is a ubiquitous approach, adapted by commonly used packages ranging from \texttt{Astroplan} \citep{MoTo2018} to \texttt{gwemopt} \citep{CoTo2018,CoAn2019,AlCo2020} for a variety of purposes.
Unfortunately, due to the inability to plan ahead, the fields chosen are not optimal; instead, an optimal schedule not only accounts for the current possible observations but also for past and potential future observations to maximize the scientific output from those observations, such as the Zwicky Transient Facility's need for a minimum 30-minute cadence when searching for transients to rule out asteroids.

Unfortunately, the scheduling problem is NP-complete and the number of observing sequences is combinatorially large.
A well-known model that can make these problems computationally tractable is the use of Integer Linear Programming (ILP); ILP problems have variables which take only discrete integer values, linear objective functions, and linear constraints. In the following, we will also use mixed ILP, which can include some non-integral variables. One popular application of this in astronomy is within the Las Cumbres Observatory (LCO)\footnote{\url{https://lco.global/}} scheduler, who operate a network of identical imagers and spectrographs; their scheduler \citep{LaSa2015} uses ILP to maximize the total number of observations obtained, weighted by the priority assigned to them by the Time Allocation Committee (TAC). ALMA solves a similar ILP model to maximize TAC-assigned scientific priorities, program completion, and telescope utilization \citep{SoMi2016}. ZTF's \blue{time-allocation scheduler uses ILP to solve both program-level and global scheduling constraints and optimally order individual observational blocks \citep{Bellm2018}, however, the schedulers used to plan within each observation block do not all use ILP, such as the greedy scheduler \texttt{gwemopt}.}
\citealt{BeFo2019}, a paper whose authorship spans many surveys and open-source astronomy software producers, advocate for community emphasis on the use of quantitative objective functions and ILP-based scheduling approaches to address the rapid proliferation of instruments, many of which will benefit from coordination.

In this paper, we introduce MUSHROOMS (Milp-Using ScHeduleR Of sky lOcalization MapS), a MILP-based scheduler for multimessenger follow-up with wide field-of-view surveys. We structure the paper as follows. We start by describing requirements faced by ZTF gravitational-wave follow-up observations in section \ref{section:requirements}. We then introduce MUSHROOMS and describe the scheduling algorithm in section \ref{section:formalism}, laying out its MILP formalism and the reasoning behind certain design decisions. In section \ref{section:simulations}, we use MUSHROOMS to schedule simulated skymaps based on LIGO-Virgo's third observing run and characterize its run time, efficiency and other relevant metrics, and summarize our results and future outlook in section \ref{section:conclusion}. 

\section{Observing Requirements}
\label{section:requirements}
\blue{Multimessenger astronomy supplements electromagnetic observations with observations using other information carriers such as gravitational waves or neutrinos.} Since 2018, ZTF has been used for target of opportunity, multimessenger follow-up searches, both searching for the sources of gravitational-wave detections during the third LIGO-VIRGO observing run \citep{CoAh2019b,AnCo2020, Kasliwal_2020}, and gamma-ray bursts from detectors like the \textit{Fermi} gamma-ray burst monitor \citep{CoAh2019,AhSi2021}.  
However, there are a number of factors one has to consider when designing schedules for such systems, both in terms of general observational requirements for ground-based surveys and certain ZTF-specific restrictions or demands.
For example, targets are only observable at night and when they are above a minimum altitude from horizontal (i.e., below a minimum airmass). 
There are also common sense constraints: for example, the scheduler cannot schedule more than one field observation at the same time, and it must restrict observations to the window of time available for observing.
In addition, there are also limits imposed by the telescope and observing system itself, such as slew speed. 

There are also a number of multimessenger transient follow-up restrictions that must be accounted for. For example, for the ZTF \blue{gravitational wave} follow-up program, there is a 30-minute cadence requirement, observing once in $r$- and $g$- bands, which serves to both eliminate asteroids and gain color information about detected transients. This requirement imposes not only a limit on the return time, but also must account for the filter exchange time within ZTF, which is 2 minutes long. Another special feature of ZTF follow-up is that the system uses a fixed grid of reference images, a preset selection of a limited number of telescope pointings to choose from.

In addition to the requirements, the goal is to limit the total amount of time required for these observations through the selection of an objective function, whose choice we will describe below.

\section{Scheduling Algorithm and MILP formulation}
\label{section:formalism}

\begin{figure}
    \centering
    \includegraphics[scale = 0.35]{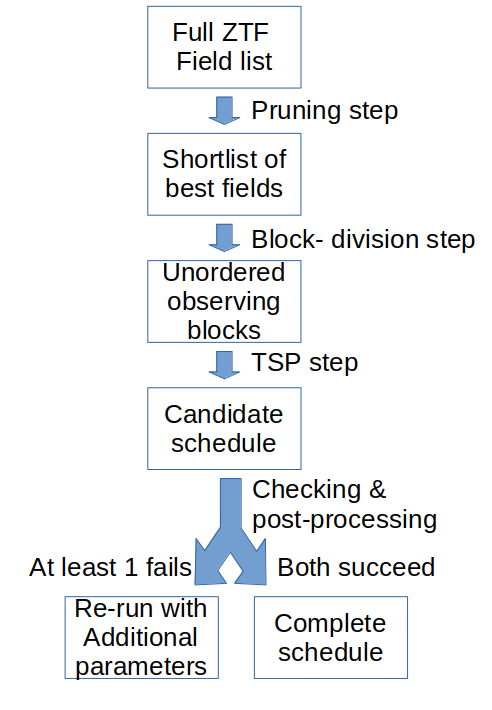}
    \caption{An illustration of MUSHROOMS' algorithm. TSP stands for travelling salesperson}
    \label{fig:algo}
\end{figure}
Due to the design of the ZTF survey and data system, ZTF has a fixed grid of 1778 telescope pointings. Given a probability density map in right ascension and declination, a span of time $t_0$ to $t_o + T$, and a fixed exposure time $\Delta t$, the goal is to produce a schedule that meets the observing requirements laid out in Sec ~\ref{section:requirements} by selecting a set of fields $S$ to observe and arranging them in time (with repeats). The objective is to maximize the total probability density contained in the area observed by at least 1 field in $S$ minus a penalty factor proportional to the amount of fields observed with proportionality constant $p$. 
\blue{We maximize the probability density contained because the overall goal of this scheduler is to identify new transients that could potentially be gravitational wave source for follow-up observation. This is done by comparing observations to reference images to find new sources, so maximizing the probability density observed in theory maximizes the probability of detecting the source for follow-up. We introduced the penalty factor $p$ with the other survey priorities in mind. It allows the user to restrict the search to only fields that introduce more than $p$ to the total probability observed. The expected range of $p$ is $[0, 0.02]$, though it has to be manually selected for each skymap. While not making it impractical for use in scheduling, using it does require some extra effort from the user to determine the trade off between observing time and detection probability best for their situation. This creates shorter duration schedules that only target the highest probability fields and are less intrusive to the other programs; with a value of zero for $p$, the schedule will fill all available time.}

\begin{figure*}
    \centering
    \includegraphics[trim = {10cm 4.5cm 10cm 6cm}, scale = 0.75]{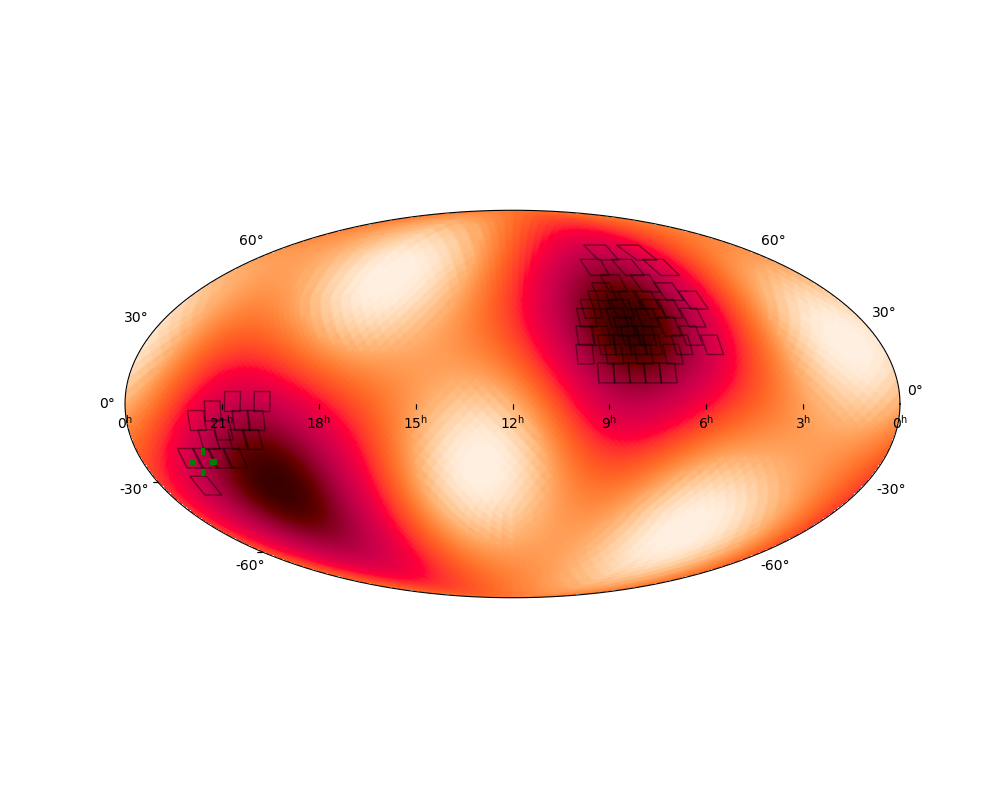}
    \caption{Skymap of simulation 119 from \citealt{PeSi2021} (see below) and the schedule created for it by MUSHROOMS. The fields selected for observation are outlined in black, and the true location of the kilonova event is highlighted with a green recticle.}
    \label{fig:skymap}
\end{figure*}

MUSHROOMS \citep{b_parazin_2022_6659827} is a python-based mixed-integer scheduler that uses the commercially available software \texttt{Gurobi}, \blue{which is free with an academic license}. When the network of ground based gravitational-wave detectors localize a new event, they release a probability map of where in the sky the source is most likely located \citep{SiPr2016}, which MUSHROOMS takes as one of its inputs. An example schedule overlaid on its corresponding probability map is shown in Fig.~\ref{fig:skymap}. For each given source localization probability map, referred to as a skymap from here on, the MUSHROOMS algorithm works in a three-step process: a preliminary pruning step, a block-division step, and an observation sequencing step. A flowchart illustrating the whole algorithm can be seen in Fig.~\ref{fig:algo}.

In the pruning step, MUSHROOMS reduces the field grid to a user-provided number of fields using a max weighted coverage algorithm \citep{nemhauser_wolsey_fisher_1978}. This is done to reduce run time during the block-division step. 

In the next step, the block-division step, a number of observing blocks are constructed out of the field shortlist produced by the pruning step. MUSHROOMS defines an observing block by a start time, an expected end time, and a collection of fields that are visible for the entire expected duration. To observe each block, all the fields within it are observed in same filter, a filter change is executed, and the fields are observed in another filter. These blocks have a minimum size which depends on the given exposure time and ensures that there are at least 30 minutes of observations between field re-observations. MUSHROOMS calculates the expected block length using an average slew time assumption of 10 seconds since the order of observations, which determines the actual slew time for each block, is not found until the next step. We use this block-division heuristic rather than giving the scheduler complete freedom to order fields and filter changes as it sees fit due to the computational complexity of complete freedom, which would require orders of magnitude more time to run. 

To minimize slew time within each block, MUSHROOMS calculates the slew times from each field within a block to all other fields in that block using a travelling salesperson (TSP) algorithm to find the order of observations that will minimize slew time within each block. 

Finally, MUSHROOMS post-processes the schedule to ensure that it is valid and satisfies all the requirements laid out in Appendix ~\ref{section:requirements}. Because the block-division algorithm uses a fixed slew time, if one of the blocks has an average slew time higher than that, the block will run longer than expected, and all subsequent blocks will have to be delayed to avoid scheduling 2 observations at the same time. There is an edge case where this delay means MUSHROOMS schedules a field for observation when it is (barely) below the visibility requirement and cannot be observed. In this scenario, we \blue{automatically} re-run the schedule utilizing a gap parameter to add more time between the offending blocks. However, in the 951 simulations utilized for this paper (see below), it did not occur once. 

The variables, constraints and objective function behind MUSHROOMS are laid out explicitly in Appendix \ref{sec:math}. This algorithm is a modification of the classic max weighted coverage problem \citep{nemhauser_wolsey_fisher_1978}, with additional constraints to allow for the creation of valid observing blocks, as well as an additional (optional) penalty factor $p$ in the objective function. 

\begin{figure}
    \centering
    \includegraphics[trim = {4cm 1cm 4cm 0.5cm}, scale = 0.35]{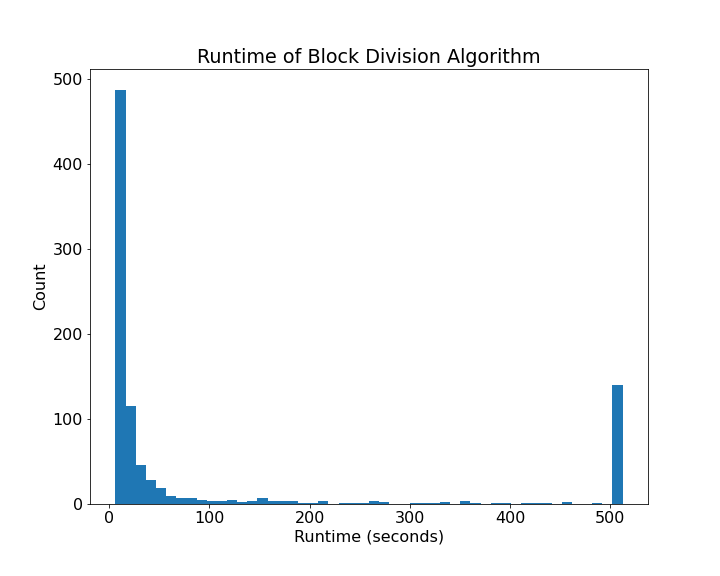}
    \caption{Run time of block-division algorithm. The large number of schedules clustered at 500 seconds is a result of setting a 500-second time limit for this step of optimization; all 500 second run times are when the MILP solver used would converge quickly on a high-quality solution but then spend the duration trying to lower the optimality gap.}
    \label{fig:runtime}
\end{figure}

\begin{deluxetable}{cllll}
    \tablecaption{\label{tab:kilonova-params}KN Light Curve Model Parameters}
    \tablecolumns{5}
    \tablewidth{0pt}
    \tablehead{
        \colhead{} &
        \multicolumn{1}{c}{Optimisic} &
        \multicolumn{1}{c}{Conservative}}
    \startdata
    Dynamical ejecta mass ($M_\odot$) & 0.005 & 0.01 \\
    Wind ejecta mass ($M_\odot$) & 0.11 & 0.01 \\
    Half opening angle & 45\arcdeg& 45\arcdeg \\
    \hline
    Peak $g$-band absolute magnitude & -15.7 & -15.1\\
    Peak $r$-band absolute magnitude & -16.0 & -15.7\\
    \enddata
\end{deluxetable}

\section{Simulated Observing Plans}
\label{section:simulations}

To assess the efficiency of the generated schedules, we ran both the MUSHROOMS and \texttt{gwemopt} algorithm on 951 simulated \blue{Binary Neutron Star detections consistent with the third LIGO-Virgo observing run (O3) from \citealt{PeSi2021}. Both \texttt{gwemopt} and MUSHROOMS were used to schedule follow-up observing plans for the 24 hours immediately following each simulated detection. Without the time to fine-tune $p$ for each skymap, MUSHROOMS was run with $p=0$ for all skymaps.} 

When performing the block-division algorithm, we recorded the run time of all 951 schedules, which can be seen in Fig.~\ref{fig:runtime}. The mean and median run times for this step were 107 and 15 seconds respectively. The large difference between the mean and median can be attributed to the 140 schedules that took the entire time limit of 500 seconds to complete. In these cases, the solver would quickly converge on a high-quality solution, but would then spend the rest of the time limit attempting to narrow the optimality gap. \blue{This 500-second time limit was chosen because when developing MUSHROOMS it was observed that most schedules which converged in a reasonable amount of time did so before 500 seconds, though it could easily be lowered to even 100 seconds without a substantial decrease in efficiency. Only an an additional 64 schedules would be truncated and use the near-optimal candidate solutions instead of a solution proven to be globally optimal.} Additionally, with a maximum number of 6 blocks (and thus a maximum of 6 filter changes in a night), the average number of filter changes scheduled was 4.7. 

\subsection{Comparisons to \texttt{gwemopt}}
\blue{For all 951 skymaps we first measured the total probability density observed by each schedule, hereafter referred to as "probability coverage." MUSHROOMS saw an average probability coverage of 0.418, while \texttt{gwemopt} had an average probability coverage of 0.387, an 8.0\% increase in probability coverage, however MUSHROOMS' schedules had an average runtime of 23700 seconds, while \texttt{gwemopt} had an average runtime of 16800 seconds, a 41.5\% increase in runtime. This is because MUSHROOMS was run with $p=0,$ meaning it filled all available time, while \texttt{gwemopt} has some logic to stop when as it gets diminishing returns by adding more fields to observe. }

\begin{figure}
    \centering
    \includegraphics[scale = 0.43]{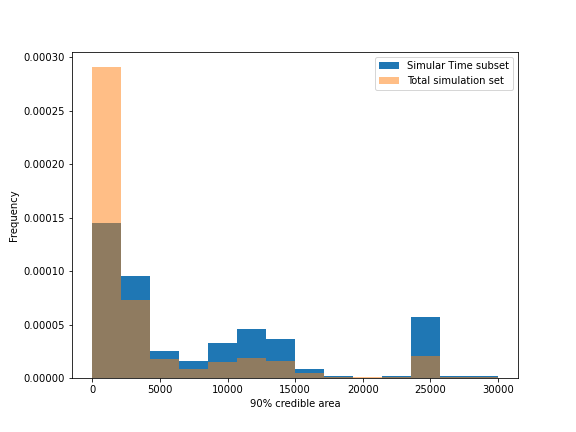}
    \caption{Frequency histograms of 90\% credible area for all 951 simulations and the 329 where MUSHROOMS and \texttt{gwemopt} produce schedules of a similar length}
    \label{fig:bias}
\end{figure}

\blue{To make a more equal comparison, we focused on the skymaps where MUSHROOMS and \texttt{gwemopt} made no more than 6 additional observations compared to the other. This value was chosen because it kept the difference in the average run time of the schedules low, while still including a large amount of skymaps. The average run times were 22900 seconds for MUSHROOMS and 22800 for \texttt{gwemopt} over this subset of 329 simulated events. An important note here is that there is selection bias here towards skymaps with a greater 90\% credible area, since they are the ones that \texttt{gwemopt} usually makes longer schedules for, as well as more well-localized schedules which, due to the event time and location, MUSHROOMS and \texttt{gwemopt} both filled almost all available time. A frequency histogram comparing the area distributions can be seen in figure \ref{fig:bias}.}

\begin{figure}
    \centering
    \includegraphics[scale=0.55]{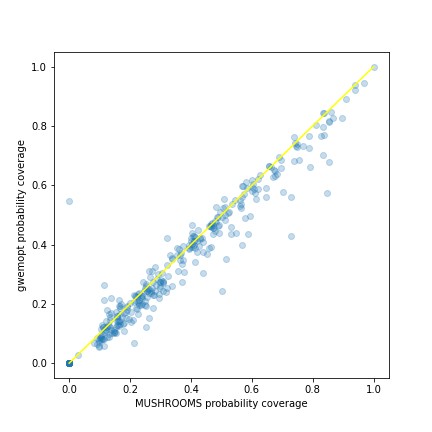}
    \caption{Probability coverage achieved by both schedulers over the similar schedule length subset. The line y=x is added in yellow for ease of comparison, with dots below the line representing schedules where MUSHROOMS out-performs \texttt{gwemopt}}
    \label{fig:pcover}
\end{figure}

\blue{For this subset, MUSHROOMS has an average probability coverage of 0.353, while \texttt{gwemopt} has an average probability coverage of 0.333, only a 5.8\% improvement for MUSHROOMS. For a skymap-by-skymap comparison, figure \ref{fig:pcover} is a scatter plot comparing the probability coverages achieved by MUSHROOMS and \texttt{gwemopt} over this subset.}

\blue{An important note, however, is that MUSHROOMS does not always out-perform \texttt{gwemopt}, even if the solution was not truncated by the 500-second time limit. This is because, even though the solution found is an optimal solution for MUSHROOMS' block-division heuristic, it may not be a globally optimal schedule. The design of MUSHROOMS forces the solutions to take on a certain format with observing blocks that are repeated in two different filters. This means the problem is (comparably) easy to implement using MILP and runs quickly, but if the best possible schedule does not fit such a format, MUSHROOMS cannot produce it. \texttt{gwemopt} has more freedom in ordering filter changes and block observations, meaning it can sometimes surpass MUSHROOMS, even with a greedy algorithm. Producing a more complex MILP formulation that lacks these restrictions and can always surpass \texttt{gwemopt} is an avenue for future research.}

\blue{For now, this means we can use a hybrid scheduler to achieve better results than either MUSHROOMS or \texttt{gwemopt} alone. By running both MUSHROOMS and \texttt{gwemopt} and using the schedule with a higher probability coverage, we can get an average coverage of 0.360, an 8.1\% improvement over \texttt{gwemopt} alone and a 2.1\% improvement over just MUSHROOMS.}

\subsection{Detection efficiency Characterizations}
\blue{Probability coverage, however, is not equivalent the actual performance a schedule will have, since it fails to capture the difficulties in identifying a kilonova, even if the field containing it is observed, a kilonova might not be detected due to it being too dim to significantly differ from the reference. As fast-fading transients, kilonova vary in magnitude significantly even over the 24 hours both schedules were allotted to search, meaning that the order of observations has a significant impact on the schedule's quality not captured by probability coverage. To address this,} following \citealt{PeSi2021}, we characterized the resulting schedules' efficiencies with \texttt{gwemopt}'s simulation and injection recovery suite for two different kilonova light curve models, \blue{this is done by injecting 10,000 kilonovae into the sky following the skymap's probability distribution, and each schedule's efficiency is the proportion of those kilonova that ZTF would have been able to detect following each schedule.}  The light curve models for the kilonovae used here, an optimistic and a conservative model, were generated by by the radiative transfer code POSSIS \citep{Bul2019} and summarized in \citealt{DiCo2020}. For details about the physical properties of each light curve, see Table \ref{tab:kilonova-params}. 

\blue{Tables \ref{tab:efficiencies_all} and \ref{tab:efficiencies_sub} compare the efficiencies of MUSHROOMS, \texttt{gwemopt} and the hybrid implemenation of the two, for all skymaps (table \ref{tab:efficiencies_all}) and for just the ones where both schedules are of a similar length (table \ref{tab:efficiencies_sub}). Due to the large number of simulations, the monte carlo uncertainty in these values is negligible. In both cases, the hybrid method out-performs both schedulers acting on their own, with an efficiency increase of about 11.5\% (11.1\%) for a conservative (optimistic) light curve in the subset where both schedules are the same length compared to just using \texttt{gwemopt} alone.}

\begin{figure*}
    \centering
    \includegraphics[scale = 0.59]{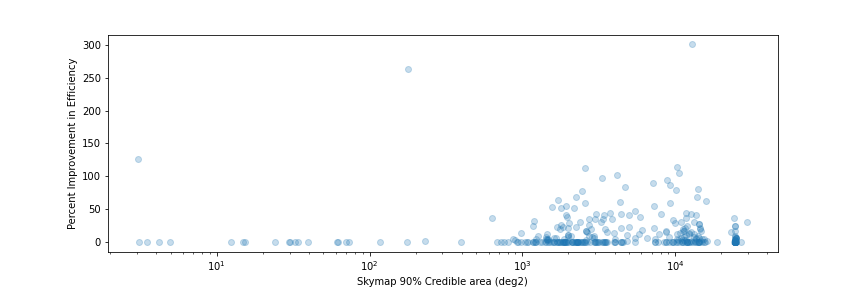}
    \caption{Percent improvement in detection efficiency by utilizing the hybrid scheduling method. A 0\% improvement means that \texttt{gwemopt} performed better than or equal to MUSHROOMS for that skymap. The three outliers at above 100\% improvement are schedules that had low efficiencies when run with \texttt{gwemopt} and a small absolute efficiency increase from MUSHROOMS resulted in a large relative increase}
    \label{fig:hybrid}
\end{figure*}

\blue{Figure \ref{fig:hybrid} compares the 90\% credible area of each skymap to the percent improvement in efficiency that would result from utilizing the hybrid method as opposed to \texttt{gwemopt} to produce a schedule for it. The selection bias against more-well localized skymaps is clear, as well as MUSHROOMS' comparative weakness at scheduling for these more localized detections. Only 37 out of 97 (38.1\%) of detections below 1000 deg$^2$ were improved upon by mushrooms, while 131 out of 232 (56.5\%) of detections above 1000 deg$^2$ were improved upon by MUSHROOMS.}

\blue{Because these smaller localizations where MUSHROOMS does worse make up a larger proportion of the total observations, this means that in actual use employing a hybrid MUSHROOMS-\texttt{gwemopt} strategy will result in less than a 11.5\% (11.1\%) improvement in detection efficiency for a conservative (optimistic) light curve. Additionally, since the hybrid strategy will never do worse than \texttt{gwemopt} alone, for a worst-case scenario, where MUSHROOMS is better for none of the remaining 622 skymaps when schedule lengths are equal, that will result in a minimum 3.1\% (3.2\%) efficiency increase for a conservative (optimistic) light curve, establishing an upper and lower bound of 11\% and 3\% respectively on the potential performance improvement of this applying hybrid method in real observing scenarios.}


\begin{deluxetable}{cllll}
    \tablecaption{\label{tab:efficiencies_all} Scheduler Efficiencies for all skymaps}
    \tablecolumns{5}
    \tablewidth{0pt}
    \tablehead{
        \colhead{} &
        \multicolumn{1}{c}{Optimistic} &
        \multicolumn{1}{c}{Conservative}}
    \startdata
    MUSHROOMS        & 0.22 & 0.21 \\
    \texttt{gwemopt} & 0.21 & 0.20 \\
    hybrid           & 0.25 & 0.23
    \enddata
\end{deluxetable}

\begin{deluxetable}{cllll}
    \tablecaption{\label{tab:efficiencies_sub} Scheduler Efficiencies for similar-length subset}
    \tablecolumns{5}
    \tablewidth{0pt}
    \tablehead{
        \colhead{} &
        \multicolumn{1}{c}{Optimistic} &
        \multicolumn{1}{c}{Conservative}}
    \startdata
    MUSHROOMS        & 0.186 & 0.163 \\
    \texttt{gwemopt} & 0.180 & 0.156 \\
    hybrid           & 0.200 & 0.174
    \enddata
\end{deluxetable}

\section{Conclusion and Outlook}
\label{section:conclusion}
In this paper, we presented a novel scheduling algorithm for scheduling wide field-of-view survey follow-up for multimessenger events, outlined its MILP formulation, and compared its performance to \texttt{gwemopt}, the \blue{Target of Opportunity} scheduler used by ZTF and other surveys in recent observing runs. 
We focused on the MUSHROOMS block-division algorithm, outlining the parameters, decision variables, objective function and constraints used to define this problem. Fundamentally, the block division algorithm is an alteration of a max weighted coverage problem, but instead of simply choosing a certain number of fields to look at, the algorithm assigns fields to variable-length blocks that are under further constraints to ensure all fields within them are observable and that no two blocks overlap. We include an additional optional penalty factor introduced into the objective function which allows for one to only observe fields that add enough probability coverage to overcome the penalty factor, leading to shorter schedules that infringe less on other programs. We also introduce a post-processing step to check for block overlaps that could be introduced by the fixed slew time approximation. 

Next, we compared MUSHROOMS to \texttt{gwemopt}, with MUSHROOMS achieving similar efficiencies \texttt{gwemopt} for both light curve models used. \blue{We showed that the differing strengths and MUSHROOMS and \texttt{gwemopt} mean when used in concert, one is able to achieve efficiencies 3\% to 11\% higher than either \texttt{gwemopt} alone.} 

The algorithm behind MUSHROOMS is a comparatively straightforward one, designed to quickly run on everyday computer hardware while still producing efficient schedules, and it already is able to increase the detection efficiency when used alongside the previous greedy scheduler. This shows the potential of using mixed-integer linear programming for scheduling multimessenger target of opportunity follow-up, and for observational scheduling as a whole, but also that there is significant room for improvement in MUSHROOMS or another Mixed-Integer scheduler, since problem formulation's rigidity in its schedules means it can still be out-done by \texttt{gwemopt} for some schedules.

There are a number of planned improvements for MILP schedulers for multimessenger follow-up. For example, MUSHROOMS does not account for the moon distance and lunar phase when scheduling observations. \blue{MUSHROOMS also does not have a straightforward way to respond to weather and other unexpected events. Currently, one would have to edit the input skymap, setting probabilities associated with affected healpix to zero before renormalizing and inputting it to MUSHROOMS. Improving it to account for both of those is important future work.} Potentially more importantly, it treats the source as having constant flux for the duration of the schedule, which is not correct for the fast transient kilonova models considered here from \citep{DiCo2020}. The most straightforward way to address this issue would be to accept a desired light curve model as an additional parameter and make an alteration to the objective function of the block division step, using that model to alter the weight of each pixel by when it is observed, such as multiplying the weight associated with that pixel by the ratio of the light curve magnitude at the observation time to the maximum magnitude. As the complete field order is not determined until the travelling salesman problem step, one may have to use an approximation of when each pixel is observed, such as the midpoint of the first block to observe a given pixel.

The block-division formulation, while a useful heuristic for limited time and computing power, has some limitations, especially when variable exposure times are desired. Producing a model that is not constrained by blocks and can jointly be optimized over the selection and ordering of all fields, subject only to the observing and time constraints, would lead to more efficient schedules. Also allowing the model to vary the exposure times of individual observations would lead to higher chances of detection because it would adjust for time and position dependent sky background. However, both improvements are much more computationally complex, and will require much greater optimization and application of high-level operations research techniques.
Using the experience gained from working on this project, among others, several authors of this paper have begun development on a more general multi-facility observation scheduling toolkit which will add those considerations into its problem formulation. 

\begin{acknowledgements}
\section{Acknowledgments}
We thank Alexander Criswell for their feedback when writing the abstract. B. Parazin acknowledges support from a Northeastern Lawrence Co-op Fellowship. 
M.~W.~Coughlin acknowledges support from the National Science Foundation with grant numbers PHY-2010970 and OAC-2117997.
S.~Anand acknowledges support from the GROWTH National Science Foundation PIRE grant 1545949.
\end{acknowledgements}

\textit{Software}: \texttt{astropy} \citep{2013A&A...558A..33A}; \texttt{matplotlib}; \texttt{ligo.skymap}\footnote{\url{lscsoft.docs.ligo.org/ligo.skymap}}


\appendix
\section{MILP Formalism}
\label{sec:math}
\subsection{Parameters}
\begin{align}
    N_{field}&\quad\textrm{Number of fields received from pruning step}\\
    N_{pix}&\quad\textrm{Number of HEALPix pixels}\\
    \{S_l\}_{l=0}^{N_{pix}-1}&\quad\textrm{Set of fields that contain pixel }l\\
    \{w_l\}_{l=0}^{N_{pix}-1}&\quad\textrm{The probability associated with HEALPix pixel }l\\ 
    \{T_{s,j}, T_{e,j}\}_{j=0}^{N_{field}-1}&\quad\textrm{The start and end observability times for field }j\\
    b_{max}&\quad\textrm{The maximum number of observation blocks to schedule}\\
    b_{size}&\quad\textrm{The minimum number of fields in an observation block}\\
    t_{exp}, t_{slew}, t_{filt}&\quad\textrm{The exposure, slew and filter change times}\\
    t_{start},t_{end}&\quad\textrm{Start and end times of the observing run}\\
    p&\quad\textrm{Penalty factor for number of fields observed}\\
    \{g_i\}_{i=0}^{i=b_{max}-2}&\quad\textrm{Additional time gap to be added between blocks i and i+1}
\end{align}
\textit{An important thing to note is that as the specific order of fields is not yet determined, $t_{slew}$ is a fixed slew time approximation. The specific slew times between each field is determined in the TSP step.}

\subsection{Binary decision variables}
\begin{align}
    \{B_{i,j}\}_{i=0,j=0}^{b_{max}-1, N_{field}-1}&\quad\textrm{Is field }j\textrm{ observed in block }i?\\
    \{y_l\}_{l=0}^{N_{pix}-1}&\quad\textrm{Is HEALPix pixel }l\textrm{ observed?}\\
    \{U_i\}_{i=0}^{b_{max}-1}&\quad\textrm{Is block }i\textrm{ used to make observations?}
\end{align}

\subsection{Continuous Decision Variables}
\begin{align}
    \{t_{o,i}\}_{i=0}^{b_{max}-1}&\quad\textrm{The starting time of block }i
\end{align}

\subsection{Objective and Constraints}
Maximize $\sum_l w_ly_l - p\sum_{i,j}B$ subject to the following constraints:

\begin{align}
\forall i, \quad \sum_{j}B_{i,j} & \geq b_{size}U_i \quad\textrm{Set minimum block size}\\
\forall i,j, \quad U_i & \geq B_{i,j} \quad\textrm{If a block makes at least 1 observation, it is being used}\\
\forall i > 0, \quad U_i & \leq U_{i-1} \quad\textrm{All unused blocks are at the end of the night}\\
\forall l, \quad \sum_{i\in S_l,j} B_{i,j} & \geq y_l \quad \textrm{A pixel is observed if it is in any observed field}\\
\forall i,j \quad t_{o,i} + 2\cdot(t_{exp} + t_{slew})\sum_{j'}B_{i,j'} + t_{filt} & \leq B_{i,j}[T_{e,j}-t_{start}-t_{exp}] + (1-B_{i,j})[t_{end}-t_{start}]\\
\textrm{A block's end}&\textrm{ time must be before the observability end time of all fields within it}\\
\forall i,j \quad t_{o,i} & \geq B_{i,j}[T_{s,j}-t_{start}]\\
\textrm{A block's start}&\textrm{ time must be after the observability start time of all fields within it}\\
\forall i > 0, \quad t_{o,i} & \geq t_{o,i-1} + 2\cdot(t_{exp} + t_{slew})\sum_{j}B_{i,j} + t_{filt} + g_{i-1}\\
\textrm{A block's start}&\textrm{ time must be after the previous block finishes}
\end{align}



\bibliographystyle{aasjournal}
\bibliography{main} 

\label{lastpage}
\end{document}